\documentclass[aip,graphicx]{revtex4-1}
\usepackage{graphicx}
\usepackage{dcolumn}
\usepackage{bm}
\usepackage{lineno}
\usepackage{physics}
\usepackage{amssymb}
\usepackage{pifont}
\usepackage[normalem]{ulem}
\usepackage{lineno}

\newcommand{\partd}[2]{\frac{\partial #1}{\partial #2}}

\setcitestyle{authoryear,round}
\begin{document}


\title{How do CPT-like symmetries shape the stability of geophysical flows?} 



\author{Tomos W. David}
\affiliation{Univ Lyon, ENS de Lyon, Univ Claude Bernard, CNRS, Laboratoire de Physique (UMR CNRS 5672), F-69342 Lyon, France}

\author{Pierre Delplace}
\affiliation{Univ Lyon, ENS de Lyon, Univ Claude Bernard, CNRS, Laboratoire de Physique (UMR CNRS 5672), F-69342 Lyon, France}

\author{Antoine Venaille}
\affiliation{Univ Lyon, ENS de Lyon, Univ Claude Bernard, CNRS, Laboratoire de Physique (UMR CNRS 5672), F-69342 Lyon, France}

\date{\today}

\begin{abstract}
We examine the role discrete symmetries, time-reversal and mirror symmetries, play in the context of geophysical waves and instabilities. By looking at three special cases from the two-layer quasi-geostrophic model as well as developing a general framework for translating real-space  transformations to Fourier-space we are able to: 1) show that baroclinic instability is an example of spontaneous parity-time symmetry breaking; 2) show that pure parity symmetry for a fluid system is exactly analogous to charge-conjugation-parity symmetry in a condensed matter system; and 3) show that when a pure parity symmetry is broken, this is associated with the suppression of wave propagation. Further, in the latter case, instability can arise without a corresponding symmetry breaking. This study highlights the role of symmetry breaking behind the dynamics of geophysical waves and instabilities.
\end{abstract}


\maketitle 

\section{Introduction}
\label{sec:introduction}
Geophysical flows are shaped by instabilities. For instance, our daily weather at mid-latitudes is largely controlled by baroclinic instability, that releases the huge amount of energy stored in planetary scale jets. A central question in geophysical fluid dynamics, and more generally in hydrodynamics, is to determine the conditions under which a given flow becomes unstable  \cite{drazin,smythcarpenter}. We will, in this study, elucidate the role that symmetry and symmetry breaking plays in the answer to this question in the context of layered models of geophysical fluids.

Flow instabilities are most often unveiled mathematically by linearizing the dynamics around a prescribed base state, and computing the spectrum of the corresponding linear operator.
The emergence of normal mode instabilities then corresponds to the appearance of imaginary eigenmodes in the spectrum, and the kinematic mechanism underlying the flow instability can usually be deduced from the polarization relation of the growing eigenmode. The generic existence of parameter regimes for which the wave spectrum is purely real may actually seem surprising as the linear wave operators encountered are in general non-Hermitian. In fact, physicists have understood over the last decades that such properties can be related to the existence of parity-time (PT) symmetries satisfied by the wave operator \cite{benderetbeottcher1998}. Those ideas have recently brought new insights to the celebrated  Kelvin-Helmholtz instability, interpreted as a case of \textit{spontaneously} broken PT-symmetry  \cite{qinetal2019, fuetal2020}. Meanwhile, the effect of other discrete symmetries that similarly constrain the spectrum of non-Hermitian operators have been found in other contexts. This is the case of the anti-PT-symmetry \cite{antonosyanetal2015,pengetal2016,yangetal2017}, also referred to as CP-symmetry in mechanical analogs of graphene, to stress the formal analogy with charge-conjugation-parity symmetry \cite{yoshidahatsugai2019,maamachekheniche2020}. 
Starting from the baroclinic instability case as a particularly insightful example, we propose here a general framework showing how PT- and CP-symmetries affect the spectrum of geophysical waves. We show what the physical manifestations are of breaking those symmetries, as revealed for instance in the structure of eigenmodes in the broken phases. 
\begin{figure}[ht!]
\centerline{\includegraphics[scale=0.9]{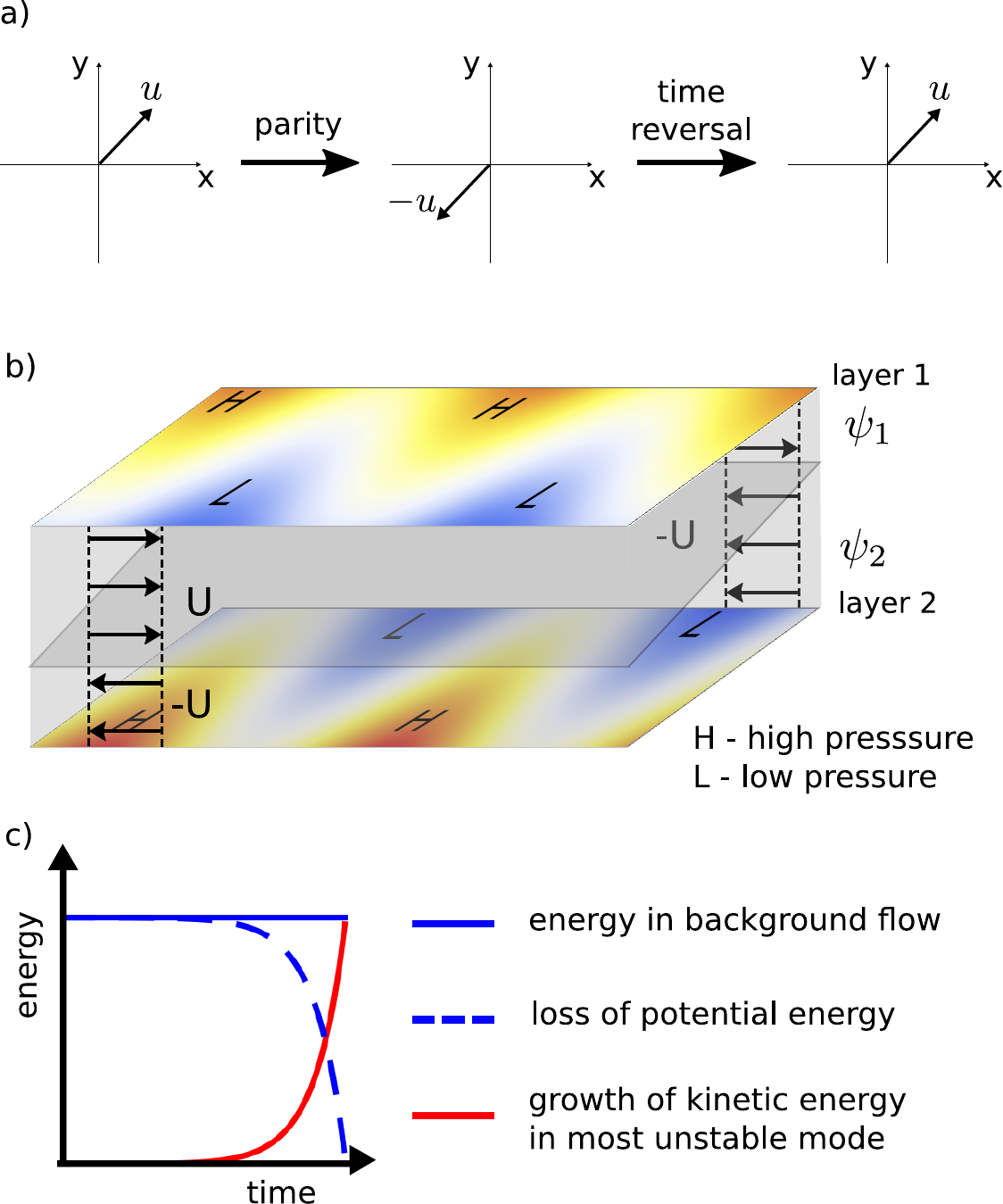}}
\caption{a) Schematic showing how velocity is symmetric with  respect to the combination of parity and time-reversal operations while being anti-symmetric with respect to each individually. We can see from this that Newton's First Law of motion, i.e. velocity is constant unless acted on by a force, is a PT-symmetric law of physics. b) Schematic of the two-layer geophysical flow model. The system is linearized about a constant but opposite background flow in each layer; on each layer the streamfunction for fastest growing mode is plotted, showing the structure of baroclinic instability. c) Demonstrates the energetics of baroclinic instability. The unstable model gains kinetic energy by converting potential energy in the large scale flow. In the linearized problems, applicable for short times, the background energy is fixed and provides an infinite reservoir of energy. In reality the large scale flow loses potential energy as shown in the schematic before non-linearity becomes important.}
\label{fig:schem}
\end{figure}

A system is PT-symmetric when its dynamics is invariant with respect to reversing the sign of all spatial and temporal coordinates, see figure \ref{fig:schem}a. The CP-symmetry we consider in this paper is understood in condensed matter physics as the combination of reversing the sign of spatial coordinates combined with exchanging particles for holes. Such a CP-symmetry was in fact recently introduced to classify non-Hermitian Hamiltonians  \cite{yoshidahatsugai2019,kawabataetal2019, delplaceetal2021}, which is  the framework we are going to use. At the level of the spectrum, this particle-hole exchange in condensed matter systems simply amounts to reversing the sign of the energy bands, which, as we shall see, has therefore a natural generalization for frequency wavebands in classical systems. The goal of this paper is both to identify the origin of these symmetries in the equations of motion, and to unveil their manifestation in the spectrum of geophysical fluid waves.

Since the PT- and CP-symmetries utilized in condensed matter physics pertain to the linear operator describing wavefunctions, it is straightforward to import them in the study of fluid waves. We will present a general framework allowing us to relate these PT- and CP-symmetries to classical time reversal symmetry, classical mirror symmetries, and their combinations, for a wide class of flow models.  We will illustrate the consequences of these symmetries for a two-layer geophysical fluid model, known as the two-layer quasi-geostrophic model (figure \ref{fig:schem}b); this model can be considered as the `fruit-fly' of geophysical fluid dynamics. 

In particular, the two-layer quasi-geostrophic model is the minimal model which admits baroclinic instability. This instability transfers energy from large-scale potential energy to smaller-scale kinetic energy in the mid-latitude atmosphere and ocean (figure \ref{fig:schem}c). We show that there is actually a variety of ways of spontaneously breaking symmetries in this model, with important physical consequences in each cases. 

In section \ref{sec:mode_merging_in_2lqg}, we introduce the linearized two-layer quasi-geostrophic model, as well as its associated wave operator in Fourier space, which we will use to exemplify the concepts discussed in this study before giving examples of modes merging under various choices of parameters. In section \ref{sec:symmetries_and_disp_rels}, we discuss the consequences of PT- and CP-symmetry on complex dispersion relations in a general context. We also show how these symmetries should be understood for a wide class of layered fluid models relevant for geophysical flows. In section \ref{sec:examples_of_symmetries}, we show that the examples discussed in section \ref{sec:mode_merging_in_2lqg} do indeed have the mathematical structure of the spontaneous PT- or CP-symmetry breaking discussed for quantum systems. In section \ref{sec:conclusions}, we summarize our results and comment on the utility of our results for the study of geophysical fluid systems.\\

\section{Merging of modes in two-layer quasi-geostrophic dynamics}
\label{sec:mode_merging_in_2lqg}

\subsection{The linearized two-layer quasi-geostrophic model}
\label{ssec:lin_qg_model}
We consider in this study a two-layer quasi-geostrophic flow model \cite{vallis}, which is the simplest flow model for mid-latitude planetary scale motion taking into account both the effect of rotation and density stratification. 
The model can be thought of as two layers of coupled two-dimensional fluids. It is often used as a simplified flow model for mid-latitude oceanic and atmospheric dynamics. A schematic of the model is shown in figure \ref{fig:schem}b.

The equations governing two-layer quasi-geostrophic dynamics are expressed as 
\begin{equation}
    \partd{q_i}{t} + \bm{u_i \cdot \nabla} q_i  = -r_i \nabla^2 \psi_i, \qquad i = 1,2,
\end{equation}
where: $r_i$ is a damping rate for linear friction in each layer; $\bm{u_i}$ is the horizontal (geostrophic) velocity on each layer;  $\psi_i$ is the streamfunction on each layer defined by its relationship to the velocity, $\bm{u_i}=\bm{\nabla} \times (\bm{k}\psi_i)$, where $\bm k$ is the vertical unit vector; and 
\begin{eqnarray}
    q_1 &=& \nabla^2 \psi_1 + \beta y + \frac{4}{L_d^2}(\psi_2 - \psi_1), \\
    q_2 &=& \nabla^2 \psi_2 + \beta y + \frac{4}{L_d^2}(\psi_1 - \psi_2),
\end{eqnarray}
are the expressions for the potential vorticity, $q_i$, on layer $i$. The potential vorticity can be thought of as the sum of: the relative vorticity $\bm{\nabla \times u_i} =  \nabla^2 \psi_i $; the planetary vorticity, $\beta y$, coming from the linear approximation to the Coriolis parameter at a latitude, $\theta$, such that $f \approx f_0(\theta) + \beta(\theta) y$; and a coupling between the layers which is associated with vortex stretching where $L_d$ is a parameter which characterizes the length scale at which effects from rotation and stratification become equally important.

We linearize these equations about a background zonal shear
\begin{eqnarray}
    \bm{u_1} &=& \bm{U} + \bm{u_1'}, \\
    \bm{u_2} &=& -\bm{U} + \bm{u_2'},
\end{eqnarray}
where $\bm U = (U, 0)$, this leads to the linearized equations (suppressing the ' notation for perturbation from here on)
\begin{eqnarray}\label{eqn:lin2lqg}
    \left(\partd{}{t} + U\partd{}{x} \right) \left[ \nabla^2 \psi_1 + \frac{4}{L_d^2} \left(\psi_2 -\psi_1 \right) \right] + \partd{\psi_1}{x} \left( \beta + \frac{8}{L_d^2} U\right) &=& -r_1\nabla^2 \psi_1, \\
    \left(\partd{}{t} - U\partd{}{x} \right) \left[ \nabla^2 \psi_2 + \frac{4}{L_d^2} \left(\psi_1 -\psi_2 \right) \right] + \partd{\psi_2}{x} \left( \beta - \frac{8}{L_d^2} U\right) &=& -r_2\nabla^2 \psi_2,
\end{eqnarray}
that are parametrized by the physical quantities $U$, $L_d$, $r_i$ and $\beta$ whose the stability of the flow depends on. We note that we consider the above problem on the infinite plane, i.e. open boundary, with constant coefficients. Consequently, the eigenmodes of the system are Fourier modes in the horizontal direction. In the language of condensed matter physics this is a a bulk problem for a stack of two-dimensional layers.

\subsection{From the classical wave operator to a matrix problem in Fourier space} 
\label{ssec:classical_to_fourier}

Our aim is to describe wave properties associated to the equations \eqref{eqn:lin2lqg}, and relate them to underlying discrete symmetries. For that  purpose, we note first that \eqref{eqn:lin2lqg} can be written formally as 
\begin{equation}
\label{eqn:general}
    \partial_t \bm \psi(x,y) = \mathcal{L}_{\bm \lambda} \left[\partial_x,\partial_y \right]  \bm \psi(x,y) 
\end{equation}
where $\bm{\psi} = (\psi_1,\psi_2)$ is a vector of the streamfunctions on each layer and $ \mathcal{L}_{\bm \lambda}$ is a $2$-by-$2$ matrix involving spatial derivatives $\partial_x$ and $\partial_y$ and the parameters
\begin{equation}
\bm \lambda = (L_d,U,\beta,r_1,r_2) .\label{eq:lambda_def}
\end{equation}
In the present case, as only time and length unit are involved in this set of parameters, there are three nondimensional parameters.
We will consider below three particular cases obtained by varying only one of the non-dimensional parameters, the others being prescribed. Having written formally the wave operator as (\ref{eqn:general}) will be useful later on to discuss symmetries.

In order to understand the waves and instabilities of these equations, it is convenient to transform equation \eqref{eqn:general} to Fourier space,  by projecting the equation onto the mode $\exp(ikx+ily-i\omega t)$, with $k$ and $l$ being the horizontal (zonal and meridional) components of the wave vector and $\omega$ is the wave frequency. This yields the eigenvalue problem 
\begin{equation}\label{eqn:tise}
    \omega \bm{\tilde{\psi}}(k,l) = \bm{H}_{\bm \lambda }[k,l]  \bm{\tilde{\psi}}(k,l), 
\end{equation}
where $\bm \lambda$ is defined in Eq. (\ref{eq:lambda_def}), 
$\bm{\tilde{\psi}} = (\tilde{\psi}_1,\tilde{\psi}_2)$ are the Fourier components of the streamfunctions and $\bm H$ is the wave operator. The dispersion relation  $\omega(k,l)$ derived from the eigenvalue problem \eqref{eqn:tise} will have the form
\begin{equation}
\label{eqn:gen_disp_rel}
    \omega_{\pm} = \frac{\tau}{2} \pm \sqrt{\Delta},
\end{equation}
where we use $\tau\equiv \mathrm{Tr} (\bm H)$ to denote the trace of $\bm H$, and the discriminant, $\Delta$, is defined as 
\begin{equation}\label{eqn:discriminant}
\Delta = \frac{\tau^2}{4} -  \mathrm{Det}(\bm H).
\end{equation}
The sign of this discriminant will play a central role to discuss the spontaneous breaking of PT- or CP-symmetry. Before addressing this question, we present below the consequences of changing the sign of this discriminant in three limiting cases for the linearized two-layer quasi-geostrophic model. 

\subsection{Frictionless case and baroclinic instability}
\label{ssec:case1_BCI}
The first choice of parameters which we will consider is the frictionless case of $r_1 = r_2 = 0$. The only nondimensional parameter to be varied is
\begin{equation}
    \gamma = \frac{\beta L_d^2}{4U}.
\end{equation}
This is the inverse of the criticality parameter involved in the classic, `textbook', case of baroclinic instability \cite[e.g.]{vallis}. The corresponding wave operator and its dispersion relation is given in appendix \ref{app:case_nor}.

We can determine whether the system is unstable by considering the sign of the contents of the square-root in \eqref{eqn:case1_disprel}, which is the discriminant of the characteristic polynomial of the wave operator. A positive discriminant here thus yields oscillatory solutions (waves) while a negative one leads to an exponentially growing mode, and thus to an instability. The region of parameter-space, $(\hat{k},\hat{l},\gamma)$, (the $\hat{}$ notation denotes dimensionless variables) for which the system is unstable is shown in Figure 
\begin{figure*}[ht!]
\centerline{\includegraphics[width=\textwidth]{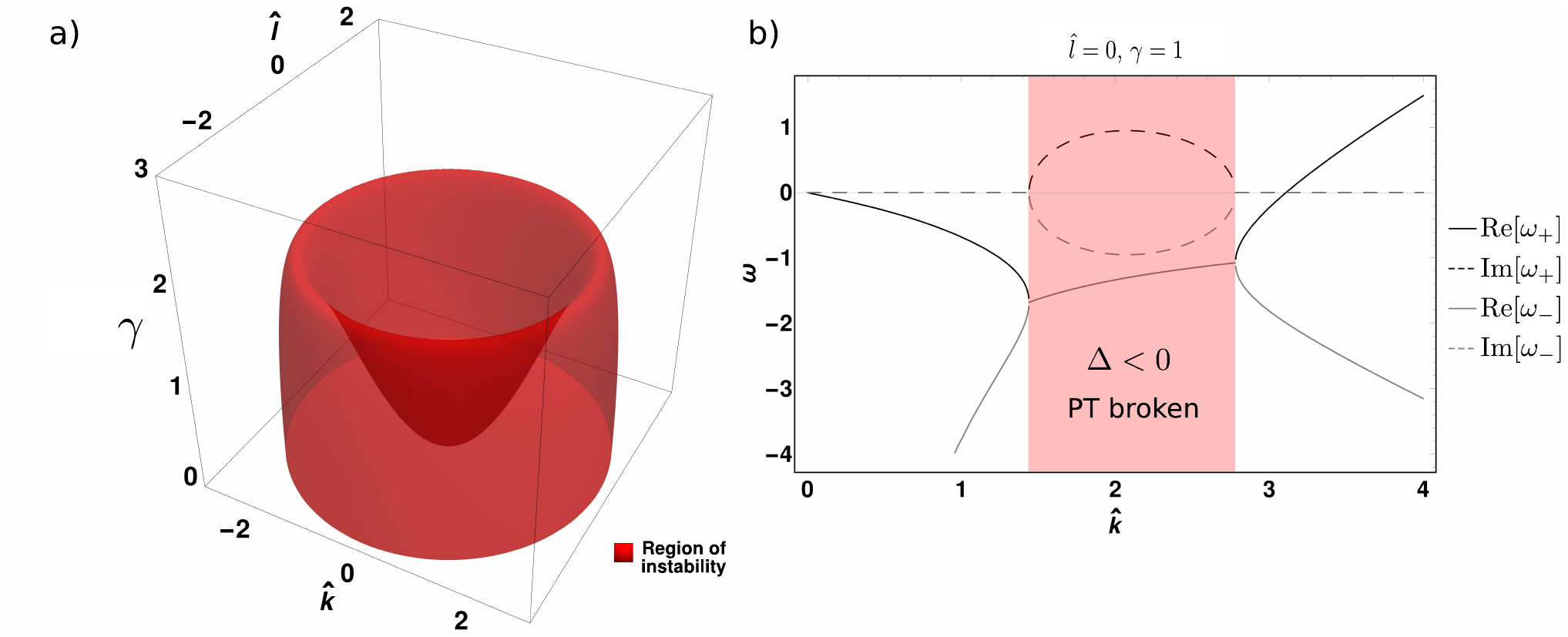}}
\caption{Behaviour of the dispersion relation for two-layer linear quasi-geostrophic model in the case of $r_1 = r_2 = 0$. a) Shows the region in the parameter space, $(\hat{k},\hat{l},\gamma)$, for which the discriminant, $\Delta$, of the characteristic polynomial is less than zero. In other words, the frequency, $\omega$, is complex and the system is unstable within the red surface. b) Gives an example of the dispersion relation where $\hat{l}=0$ and $\gamma = 1$. We see the real parts if the two modes merge at the point when the imaginary parts of the modes become non-zero and the system becomes unstable. These points where the modes merge are known as exceptional points, where the real and imaginary parts of the a number of modes become equal. The red surface in a) defines the set of all exceptional points as the parameters are varied.}
\label{fig:nor_reg_disp}
\end{figure*}
\ref{fig:nor_reg_disp}a as the volume enclosed by the red surface. We note that the surface of the critical points (transition from stable to unstable) is isotropic in the $(\hat{k},\hat{l})$ plane, which reflects that fact that the discriminant, $\Delta$, depends only on $\hat{K}^2 = \hat{k}^2 + \hat{l}^2$ in that case.

Figure \ref{fig:nor_reg_disp} shows an example for the parameters, $\hat{l}=0$ and $\gamma = 1$. We see an unstable region bounded by two points where the two modes of the system have merged. At these points, the real and imaginary parts of each frequency are equal; these points are known as \textit{exceptional points} and have recently attracted a lot of attention in non-Hermitian topological photonics and condensed matter physics \cite{bergholtzetal2021}.

\subsection{Stability in the case of friction and anti-friction}
\label{ssec:case2_FAF}

The second choice of parameters which we will consider describes a situation without mean flow $U = 0$ and an opposite friction in each layer $r_1=-r_2=r$. This choice of parameters may seem `exotic' from the perspective of relevance to geophysical flows. It corresponds to a case with gain and loss in the upper and lower layer, respectively. We include it here because we see the same behaviour of merging modes that we observed in the classic baroclinic instability discussed above. The only nondimensional parameter of the problem is then \begin{equation}
    \delta = \frac{L_d \beta}{r}.
\end{equation}

 The wave operator and the dispersion relation corresponding to this particular case are given in appendix \ref{app:case_noU}. 

\begin{figure*}[ht!]
\centerline{\includegraphics[width=\textwidth]{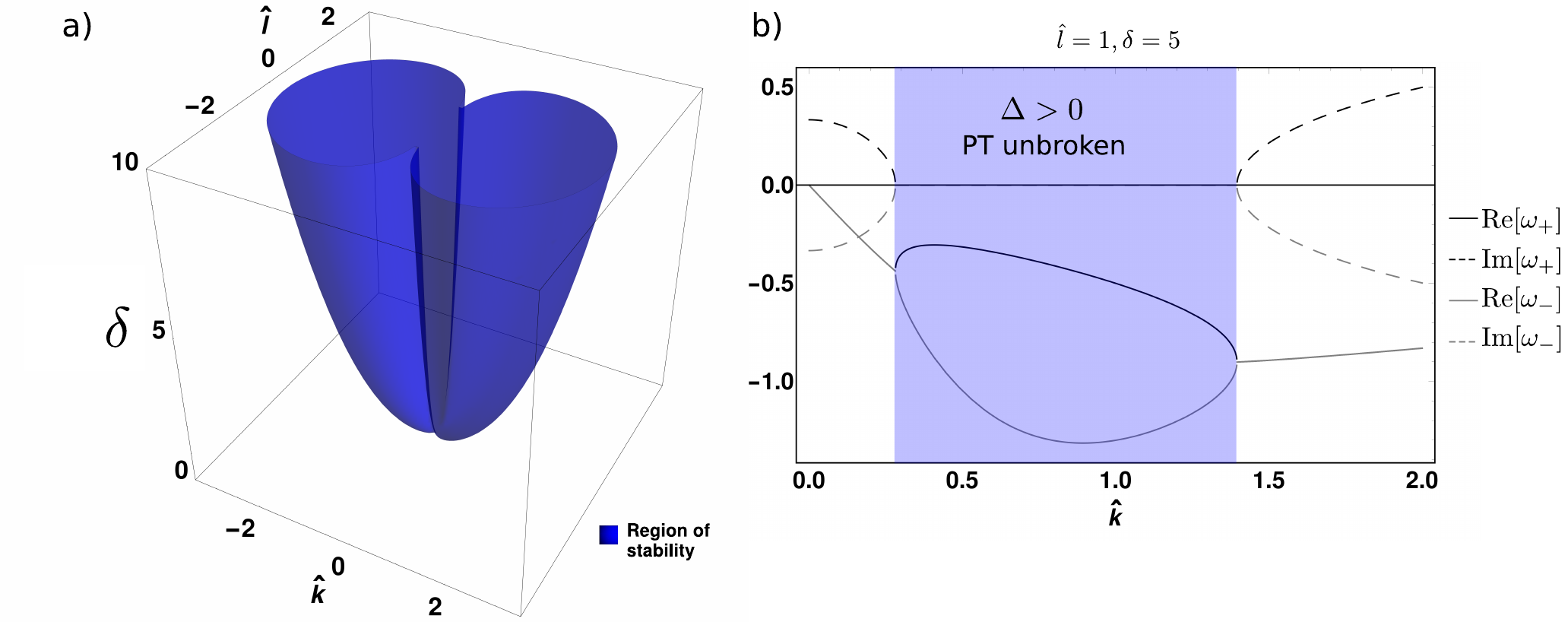}}
\caption{Behaviour of the dispersion relation for two-layer linear quasi-geostrophic model in the case of $U=0$ and $r_1=-r_2=r$. a) Shows the region in the parameter space, $(\hat{k},\hat{l},\delta)$, for which the discriminant of the characteristic polynomial is greater than zero. In other words, the frequency, $\omega$, is real and the system is stable within the blue surface. b) Gives an example of the dispersion relation where $\hat{l}=1$ and $\delta = 5$. We see the real parts if the two modes merge at the point when the imaginary parts of the modes become non-zero and the system becomes unstable. These points where the modes merge are known as exceptional points, where the real and imaginary parts of the a number of modes become equal. The blue surface in a) defines the set of all exceptional points are the parameters are varies. }
\label{fig:noU_reg_disp}
\end{figure*}

Once again we find that the  discriminant, $\Delta$, has a well-defined sign, so that we can determine whether the system is unstable by looking at where it is negative.
We describe in figure \ref{fig:noU_reg_disp}  a transition from unstable to stable states, which are located inside the blue surface. In figure \ref{fig:noU_reg_disp}b we see the same mode merging behaviour as in figure \ref{fig:nor_reg_disp}b. 

\subsection{Suppression of wave propagation}
\label{ssec:case3_SWP}

The third choice of parameters which we will consider describes a situation without planetary vorticity gradients, $\beta = 0$, and a similar friction in each layer $r_1 = r_2 = r$. The only nondimensional parameter of the problem is then
\begin{equation}
    \alpha = \frac{L_d r}{U}. 
\end{equation}
The corresponding wave operator and dispersion relations are given in appendix \ref{app:case_nob}. When only bottom friction is present, the parameter $1/\delta$ is  sometimes called throughput parameter, and is known to play a key role in shaping the flow structures of quasi-geostrophic turbulence \cite{arbic,ribbon,gallet2020vortex}. 

\begin{figure*}[ht!]
\centerline{\includegraphics[width=\textwidth]{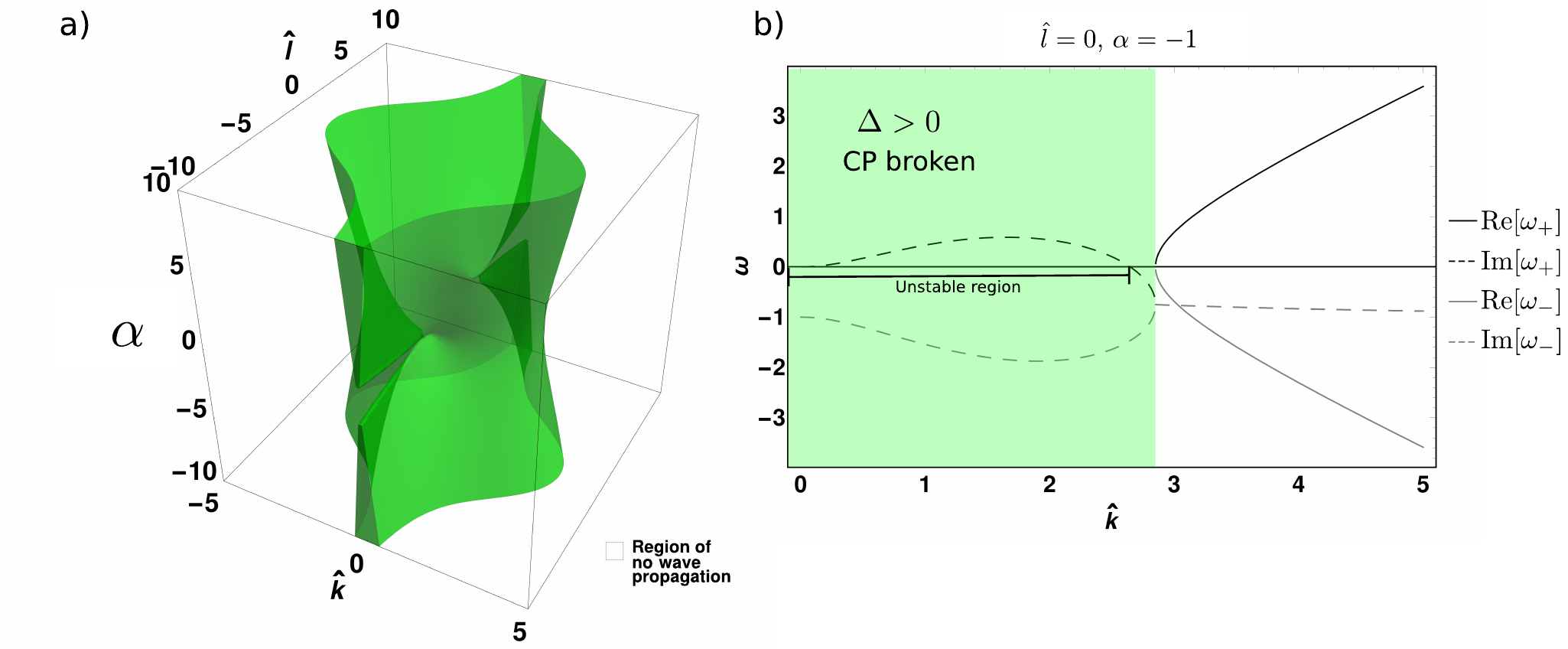}}
\caption{Behaviour of the dispersion relation for two-layer linear quasi-geostrophic model in the case of $\beta=0$ and $r_1=r_2=r$. a) Shows the region in the parameter space, $(\hat{k},\hat{l},\alpha)$, for which the discriminant of the characteristic polynomial is greater than zero. In this case, the frequency, $\omega$, is purely imaginary and the system is cannot have propagating waves within the green surface. b) Gives an example of the dispersion relation where $\hat{l}=0$ and $\alpha = -1$. We see the imaginary parts if the two modes merge at the point when the real parts of the modes become non-zero and the system admits propagating waves. We note that for this choice of parameters the system does become unstable (see b)) but this is unrelated to the merging of modes. These points where the modes merge are known as exceptional points, where the real and imaginary parts of the a number of modes become equal. The blue surface in a) defines the set of all exceptional points are the parameters are varies. }
\label{fig:nob_reg_disp}
\end{figure*}

In this case, the behaviour of the dispersion relation is different to the cases we previously discussed. Although the discriminant, $\Delta$, still has a well defined sign, the introduction of an imaginary trace, see equation \eqref{eqn:gen_disp_rel}, changes the physics of mode merging. When $\Delta$ is negative, the frequency becomes purely imaginary, and, since the real part is zero, we cannot have propagating waves in the region of parameter-space.  This region is shown contained within the green surface of  figure \ref{fig:nob_reg_disp}a.
 Additionally, we realize that because of the imaginary first term of the dispersion relation the imaginary part of the two frequencies are not the negative of each other, as was the case in \ref{ssec:case1_BCI} and \ref{ssec:case2_FAF}. This means that in this case we cannot associate the merging of modes (exceptional points) with the transition between stable and unstable regimes. An example of this is given in figure \ref{fig:nob_reg_disp}, $\hat{l}=0$ and $\alpha = -1$, where the system becomes unstable for a lower value $\hat{k}$ than the point where the modes merge. We note here that although the physics of mode merging has changed (describing the non-propagation of wave), the system can nevertheless become unstable as marked in figure \ref{fig:nob_reg_disp}.

\section{Symmetries and their consequences for dispersion relations}
\label{sec:symmetries_and_disp_rels}

In order to explain the three different regimes of mode merging described above, we build upon the work by \cite{qinetal2019,fuetal2020} who related the emergence of Kelvin-Helmholtz instability to the phenomenon of spontaneous PT-symmetry breaking. We need for that purpose to introduce the general condensed matter physics framework for PT- and CP-symmetries.

The wave dynamics is described in Fourier space by an eigenvalue problem of some wave operator, $\bm{H_\lambda}$,  which is a function of the horizontal wavenumbers and $\bm \lambda$ denotes the set of physical parameters. Such an eigenvalue problem (see Eq. \eqref{eqn:tise}) can be interpreted as a time independent Schr\"{o}dinger equation, that allows us to use the notion of PT-symmetry, as introduced below, in analogy with quantum mechanics Hamiltonians.
Ideas around PT-symmetry have also been developed in the realm of classical Hamiltonians \cite{bender}. Note that the wave operator we use here is not the Hamiltonian of the classical system; it describes its frequency spectrum  which is not related to the energy unlike in the case of quantum mechanics.

\subsection{PT- and CP-symmetries}
\label{ssec:def_of_PTandCP}

PT-symmetry is an ubiquitous symmetry, encountered in various contexts from particles physics to classical mechanics. In quantum physics, it provides a non-Hermitian extension of Hamiltonians, such that the energy spectrum may remain real valued. Formally, a quantum system is said to be PT-symmetric if there exists a unitary operator $\bm{U}$ such that its Hamiltonian $\bm{H_\lambda}$ satisfies
\begin{equation}\label{eqn:PT_symm_def}
    \bm{U H^*_\lambda U^{-1}} = \bm{H_\lambda}
\end{equation}
where $*$ stands for the complex conjugation, and $\bm{\lambda}$ is a set of parameters. We can also state that the Hamiltonian commutes with an anti-unitary operator.

In our case, this Hamiltonian is replaced by the wave operator introduced in Eq. \eqref{eqn:tise}. In the quantum context, $\bm{\lambda}$ refers to the momentum, and the complex conjugation stems from the anti-unitary property of  time-reversal symmetry in quantum mechanics, hence the invariance of the system under the combination of parity (that reverses momentum) and time-reversal (that also reverses momentum), denoted by PT-symmetry when the equation \eqref{eqn:PT_symm_def} holds.  

Now we consider the effect of PT-symmetry, using \eqref{eqn:PT_symm_def}, on the dispersion relation given by the eigenvalues of $\bm{H_\lambda}$. For a PT-symmetric system we can, hence, write that
\begin{equation}
    \mathrm{det}(\bm{H_\lambda} - \omega) = \mathrm{det}(\bm{H^*_\lambda} - \omega),
\end{equation}
which, for a $2$-by-$2$ matrix will lead to the following equality between characteristic quadratics
\begin{equation}
    \omega^2 - \tau \omega + d = \omega^2 - \tau^* \omega + d^*
\end{equation}
where we recall that $\tau = \mathrm{Tr}(\bm{H_{ \lambda}})$ and with $d = \mathrm{det}(\bm{H_{ \lambda}})$. This implies that for a PT-symmetric two dimensional system, the trace and determinant will be real. The discriminant of the characteristic quadratic of $\bm{H_{ \lambda}}$, $\Delta$, will also be real. As a consequence, when $\Delta > 0$, $\mathrm{Im}(\omega_\pm) = 0$ so that we have two stable wave-like solutions, while when $\Delta <0$, $\omega_+ = \omega_-^*$ and thus  the flow is unstable. The critical point $\Delta = 0$ corresponds to an exceptional point. This behaviour is exactly that of the dispersion relations  that we discussed in sections \ref{ssec:case1_BCI} and \ref{ssec:case2_FAF}.

Another discrete symmetry which bears resemblance to PT-symmetry is that of CP-symmetry (or anti-PT-symmetry), that will turn out to have interesting relevance to geophysical fluids.
At the level of the wave operator, CP-symmetry is formally defined as 
\begin{equation}\label{eqn:CP_symm_def}
  \bm{U H^*_\lambda U^{-1}} = -\bm{H_\lambda}
\end{equation}
 where $\bm U$ is some unitary operator and $\bm{\lambda}$ is some set of parameters on which a Hamiltonian may depend. We can also state that the Hamiltonian anti-commutes with an anti-unitary operator. For a CP-symmetric system we can, hence, write that
\begin{equation}
    \mathrm{det}(\bm{H(\lambda)} - \omega) = \mathrm{det}(\bm{-H^*(\lambda)} - \omega),
\end{equation}
leading to the characteristic quadratic
\begin{equation}
    \omega^2 - \tau \omega + d = \omega^2 + \tau^* \omega + d^*,
\end{equation}
from which it is inferred that $\tau$ must be imaginary and $d$ must be real again resulting in a real discriminant, $\Delta$. Therefore, when $\Delta> 0$, then $\mathrm{Re}(\omega_\pm) \neq 0$ and there are thus wave-like solutions. But since $\mathrm{Im}(\omega_\pm)$ is equal for both modes, these waves will be either unstable or decaying depending on the sign of $\mathrm{Im}(\omega_\pm)$. In contrast, when $\Delta < 0$, we have  $\mathrm{Re}(\omega_\pm) = 0$ meaning that propagating waves are suppressed. Similarly to the PT-symmetric case, an exceptional point appears at the transition between those two regimes, at $\Delta$ = 0.
Those generic behaviours are exactly those observed in the dispersion relations of frictional two-layer model with no rotation discussed in section \ref{ssec:case3_SWP}.

Both of the classes of symmetries discussed above can be \emph{spontaneously broken}; this is where the eigenvectors of the Hamiltonian or wave operator cease to be the mutual eigenvectors of the anti-unitary operator associated with the symmetry. In the case of PT-symmetry, spontaneous symmetry breaking occurs when $\Delta < 0$. In the case of CP-symmetry, spontaneous symmetry breaking occurs when $\Delta > 0$.

We have seen that the two-layer quasi-geostrophic model exhibits behaviour in its complex dispersion relation which is indicative of either PT-symmetry  (sections \ref{ssec:case1_BCI} and \ref{ssec:case2_FAF}) or CP-symmetry (section \ref{ssec:case3_SWP}). We explain below how to actually identify PT- and CP- symmetries from the direct inspection of a wide class of flow models. 

\subsection{Symmetries of a general layered fluid model}
\label{ssec:general}

\subsubsection{Real-space symmetries}
\label{sssec:real_space_symm}
Here we will discuss the effect of discrete transformations on the operator $\mathcal{L}_{\bm{\lambda}}$ which appears in the general layered fluid model given in \eqref{eqn:general}. We keep the fields (the components of the vector $\bm \psi$) very general. They could be, for example,
the streamfunction on each layer in the case of quasi-geostrophic models or, alternatively, the
velocity components plus the layer interface displacement in the case of shallow water models. Many geophysical fluid models can be recast in the form given
by \eqref{eqn:general}, see e.g. \cite{onuki2020} . We are interested here in representing the symmetries of such models in real and Fourier spaces.

The basic discrete transformations are 
\begin{eqnarray}
\text{time-reversal, }\,    & T &:\, \{x,y,t,\bm \psi\} \mapsto \{x,y,-t,\bm{R_t  \psi}\}, \\
\text{reflection in $x$, }\,      & P_x &:\, \{x,y,t,\bm \psi\} \mapsto \{-x,y,t,\bm{R_x  \psi}\}, \\
\text{reflection in $y$, }\,     & P_y &:\, \{x,y,t,\bm \psi\} \mapsto \{x,-y,t,\bm{R_y \psi}\}, \\
\text{reflection in $z$, }\,     & P_z &:\, \{x,y,t,\bm \psi\} \mapsto \{x,y,t,\bm{R_z \psi}\}, 
\end{eqnarray}
where $\bm{R_t}$, $\bm{R_x}$ and $\bm{R_y}$ are  diagonal matrices which act on vectors $\bm \psi$ by changing the sign of each dynamical variable which is physically consistent with the transformation. For example, if the fields of interest are the velocity components, $\bm \psi = (u,v)$, then they will become $(-u,-v)$ under time-reversal and hence, $\bm{R_t} = \textrm{diag}(-1,-1)$. Since the flow models involves stacked layers in the $z$ direction, the matrix $\bm{R_z}$ is an operator which reflects the variables in the $z=0$ plane. For instance, in a two-layer quasi-geostrophic model $\bm \psi = (\psi_1,\psi_2) \mapsto (\psi_2, \psi_1)$ under a mirror transformation in the vertical direction and hence, $\bm{ R_z} = \begin{pmatrix}0 & 1 \\ 1 & 0 \end{pmatrix}$. 

In order to understand the  symmetries of $\mathcal{L}$, we need to transfer the action of these transformations from the coordinates and fields onto the operator $\mathcal{L}$. By inspecting equation \eqref{eqn:general} we find the equivalent transformations
\begin{eqnarray}
    & T &:\, \mathcal{L}_{\bm \lambda} \left[\partial_x,\partial_y \right] \mapsto -\bm{R_t}\mathcal{L}_{\bm \lambda} \left[\partial_x,\partial_y \right] \bm{R_t^{-1}},\\
    & P_x &:\, \mathcal{L}_{\bm \lambda} \left[\partial_x,\partial_y \right] \mapsto \bm{R_x}\mathcal{L}_{\bm \lambda} \left[-\partial_x,\partial_y \right]\bm{R_x^{-1}},\\
    & P_y &:\, \mathcal{L}_{\bm \lambda} \left[\partial_x,\partial_y \right] \mapsto \bm{R_y}\mathcal{L}_{\bm \lambda} \left[\partial_x,-\partial_y \right]\bm{R_y^{-1}},\\
    & P_z &:\, \mathcal{L}_{\bm \lambda} \left[\partial_x,\partial_y \right] \mapsto \bm{R_z}\mathcal{L}_{\bm \lambda} \left[\partial_x,\partial_y \right] \bm{ R_z^{-1}}.
\end{eqnarray}

In general, the system will not be symmetric under these basic transformations but might be symmetric with respect to combinations of these transformations. As will shall see, such combinations are generally needed to find fluid analogues of `parity-time' or `charge-conjugation-parity' symmetries found in quantum physics. 

\subsubsection{Fourier-space symmetries}
\label{sssec:fourier_space_symm}
Now that we have established how the operator $\mathcal{L}$ transforms under mirroring and time-reversal, we now show how these transformations look in Fourier space. By using the ansatz $\psi = \hat{\psi} \exp(i(\bm{k\cdot x} - \omega t ))$, with $\bm k = (k,l)$, equation \eqref{eqn:general} becomes 
\begin{equation}
    \omega \bm{\hat{\psi}}(k,l) = \bm{H}_{\bm \lambda}[k,l]  \bm{\hat{\psi}}(k,l).
\end{equation}
By comparing with the transformation in real-space we can write that the wave operator transforms as
\begin{eqnarray}
\label{eq:def_sym}
    & T &:\, \bm{H_\lambda} \left[k,l \right] \mapsto -\bm{R_t}\bm{H_\lambda} \left[k,l \right] \bm{R_t^{-1}},\\
    & P_x &:\, \bm{H_\lambda} \left[k,l \right] \mapsto \bm{R_x}\bm{H_\lambda} \left[-k,l \right]\bm{R_x^{-1}},\\
    & P_y &:\, \bm{H_\lambda} \left[k,l \right] \mapsto \bm{R_y}\bm{H_\lambda} \left[k,-l \right]\bm{R_y^{-1}},\\
    & P_z &:\, \bm{H_\lambda} \left[k,l \right] \mapsto \bm{R_z}\bm{H_\lambda} \left[k,l \right] \bm{ R_z^{-1}}.
\end{eqnarray}
It is useful at this stage to combine the horizontal mirror transformations into a two-dimensional parity transformation, $P_{xy} = P_xP_y$, defined as
\begin{equation}
    P_{xy}: \, \bm{H_\lambda} \left[k,l \right] \mapsto \bm{R_{xy}}\bm{H_\lambda} \left[-k,-l \right]\bm{R^{-1}_{xy}},
\end{equation}
This combination is useful because of the consequence of changing the sign of both wavenumbers at the same time for the wave operator; it can be shown, see appendix \ref{app:proof}, that 
\begin{equation}\label{eqn:key_condition}
    \bm{H_\lambda} \left[-k,-l \right] = -\bm{H^*_\lambda} \left[k,l \right].
\end{equation}
This comes from the fact that the elements of the wave operator, $\bm{H_\lambda}$, must be polynomial functions of $ik$ and $il$ and that the original partical differential equations have real coefficients. 

Thus we can rewrite the two-dimensional parity operator, $P_{xy} = P_xP_y$, and write its action on the wave operator as
\begin{equation}
    P_{xy}: \, \bm{H_\lambda} \left[k,l \right] \mapsto -\bm{R_{xy}}\bm{H^*_\lambda} \left[k,l \right]\bm{R^{-1}_{xy}},
\end{equation}
where we have defined $\bm{R_{xy}} = \bm{R_xR_y}$. By combining the horizontal mirror transformations and knowing the relation \eqref{eqn:key_condition} we see that having $P_{xy}$ symmetry or, a symmetry consisting of $P_{xy}$ in combination with one or more of the other transformations, lead to the introduction of the complex conjugate thus introducing the anti-unitary structure associated with the definitions of PT- or CP-symmetry which were discussed in section \ref{ssec:def_of_PTandCP}. Here we have also proved that if a system of partial differential equations of the form \eqref{eqn:general} is found to be symmetric by inspection with respect to some combination of the transformations discussed above, then its dispersion relation can exhibit mode-merging and the characteristic behaviour of spontaneous PT- or CP-symmetry breaking. This can be inferred without making the transformation of the equations into Fourier space as we will see in section \ref{sec:examples_of_symmetries}.

Having written the action of the basic transformations on the wave operator allows us to
establish the dictionary between those, that act on the wave operator $\bm{H_\lambda} \left[k,l \right]$ in Fourier space, and the discrete space-time symmetries of the linearized system governed by $\mathcal{L}_{\bm \lambda}[\partial_x, \partial_y,]$, as summarized in table \ref{table}.

\begin{table}[h!]
    \centering
    \begin{tabular}{l|c}
         Symmetry of the linearized flow-model & Symmetry in Fourier space   \\
         \hline
        i)\quad 2D parity , $P_{xy}$       &   CP \\
       ii)\quad  2D parity-time , $P_{xy}T$ &  PT  \\
       iii)\quad  3D parity , $P_{xy}P_z$  &  CP   \\
       iv)\quad  3D parity-time , $P_{xy}P_zT$ &   PT
    \end{tabular}
    \caption{Formal interpretation of the parity and time symmetries of the linearized layered  flow model of the form \eqref{eqn:general} in two and three spatial dimensions, as emergent anti-unitary PT and CP symmetries for the wave operator in Fourier space $\bm{H_\lambda} \left[k,l \right]$ .}
    \label{table}
\end{table}

In the following section, 
we will show that cases ii), iii) and iv) are exemplified by the three special cases of the two-layer quasi-geostrophic model which we discussed in sections \ref{ssec:case1_BCI}, \ref{ssec:case2_FAF} and \ref{ssec:case3_SWP}. Case i) can for instance be applied to the shallow water model with friction, allowing us to highlight the underlying key role of the in-plane inversion symmetry $P_{xy}$ to explain the emergence of the CP symmetry in the wave operator found in  \cite{delplaceetal2021}.

\section{Examples of PT- and CP-symmetry breaking in geophysical fluids}
\label{sec:examples_of_symmetries}
\subsection{Symmetries of the two-layer quasi-geostrophic equations}
\label{ssec:symm_2lqg}
Having established the relationship between the definition of mirror and time-reversal transformations in coordinate space with the anti-unitary operators which act on the wave operator in Fourier-space, we are now able to unveil the PT and CP symmetries of layered linear fluid models. We will now do so for the two-layer quasi-geostrophic examples we discussed (equations \eqref{eqn:lin2lqg}). In that case, the transformations \eqref{eq:def_sym} become
\begin{eqnarray}
    & T &: \{x,y,t,\psi_1,\psi_2\} \mapsto \{x,y,-t,-\psi_1,-\psi_2\},\\
    & P_{xy} &: \{x,y,t,\psi_1,\psi_2\} \mapsto \{-x,-y,t,\psi_1,\psi_2\}, \\
    & P_z &: \{x,y,t,\psi_1,\psi_2\} \mapsto \{x,y,t,\psi_2,\psi_1\} \\
\end{eqnarray}
so we can immediately write that
\begin{equation}
    \bm{R_t} =
    \begin{pmatrix}
    -1 & 0  \\
    0 & -1 \\
    \end{pmatrix}, \,
    \bm{R_{xy}} =
    \begin{pmatrix}
    1 & 0  \\
    0 & 1 \\
    \end{pmatrix}, \,
    \bm{R_z} =
    \begin{pmatrix}
    0 & 1  \\
    1 & 0 \\
    \end{pmatrix}.
\end{equation}
Next we express the relevant  PT or CP symmetries for each of the cases we discussed in sections \ref{ssec:case1_BCI}, \ref{ssec:case2_FAF} and \ref{ssec:case3_SWP} by writing its unitary parts that are obtained as specific combinations of $\bm{R_t}$, $\bm{R_{xy}}$ and $\bm{R_z}$. 

\subsubsection{Frictionless baroclinic case}
\label{sssec:ex_symm_case1}

In the case where $r_1=r_2=0$, we find that the system has a $P_{xy}T$ symmetry, reflection in the horizontal spatial dimensions combined with time-reversal, whose corresponding operator thus reads 
\begin{equation}
    \bm{R_{xy}R_t} =
    \begin{pmatrix}
    -1 & 0  \\
    0 & -1 \\
    \end{pmatrix} 
\end{equation}
which is basically identity. The Fourier space representation of the $P_{xy}T$-transformation operator, that acts on the wave operator, is then  obtained by combining this unitary part with complex conjugation, as defined in section \ref{ssec:def_of_PTandCP}, that is $\bm{U_{1}} \kappa$. The behavior of the eigenvalues of the wave operator, that dictates the stable and unstable regimes, as shown in figure \ref{fig:nob_reg_disp}, thus follows from this  symmetry.

Moreover, beyond the complex nature of the frequency spectrum, the $P_{xy}T$ symmetry also constrains the structure of the eigenmodes specifically in the stable and unstable regimes. To see it, let us consider a generic $2\times 2$ wave operator that satisfies this PT symmetry. Its matrix representation must therefore satisfy
\begin{equation}
    \begin{pmatrix}
    a & b \\
    c & d 
    \end{pmatrix} 
    =
    \begin{pmatrix}
    a^* & b^* \\
    c^* & d^*
    \end{pmatrix}
\end{equation}
and thus all the coefficients are real-valued. 
The eigenvectors read
\begin{equation}
    \bm{\psi_{\pm}} = 
    \begin{pmatrix}
    b \\ 
    \frac{\tau}{2} - a \pm \frac{\sqrt{\Delta}}{2}
    \end{pmatrix} \ .
\end{equation}
so that they remain real valued as long as $\Delta>0$. In that case 
\begin{equation}
    \bm{R_{xy}R_t}\kappa \bm{\psi_{\pm}} = -\bm{\psi_{\pm}^*} = -\bm{\psi_{\pm}},
\end{equation}
meaning that an eigenstate of the wave operator remains an eigenstate of the PT symmetry operator $\bm{R_{xy}R_t}\kappa$.
In contrast, when $\Delta < 0$, it follows that 
\begin{equation}
    \bm{R_{xy}R_t}\kappa \bm{\psi_{\pm}} = -\bm{\psi_{\pm}^*} = -\bm{\psi_{\mp}},
\end{equation}
since the only part of the eigenvector that is changed under complex conjugation is, $\pm i|\sqrt{\Delta}| \mapsto \mp i|\sqrt{\Delta}|$. The eigenvectors of the wave operator are not eigenvectors of the $P_{xy}T$ symmetric operator, while those two operators still commute. This is known as the \textit{spontaneously breaking} of the PT-symmetry. It follows that the baroclinic instability, as described by the two-layer model, is associated to a $P_{xy}T$ symmetry breaking, where $P_{xy}$ is the in-plane  inversion symmetry. 

\begin{figure}[ht!]
\centerline{\includegraphics[scale=0.9]{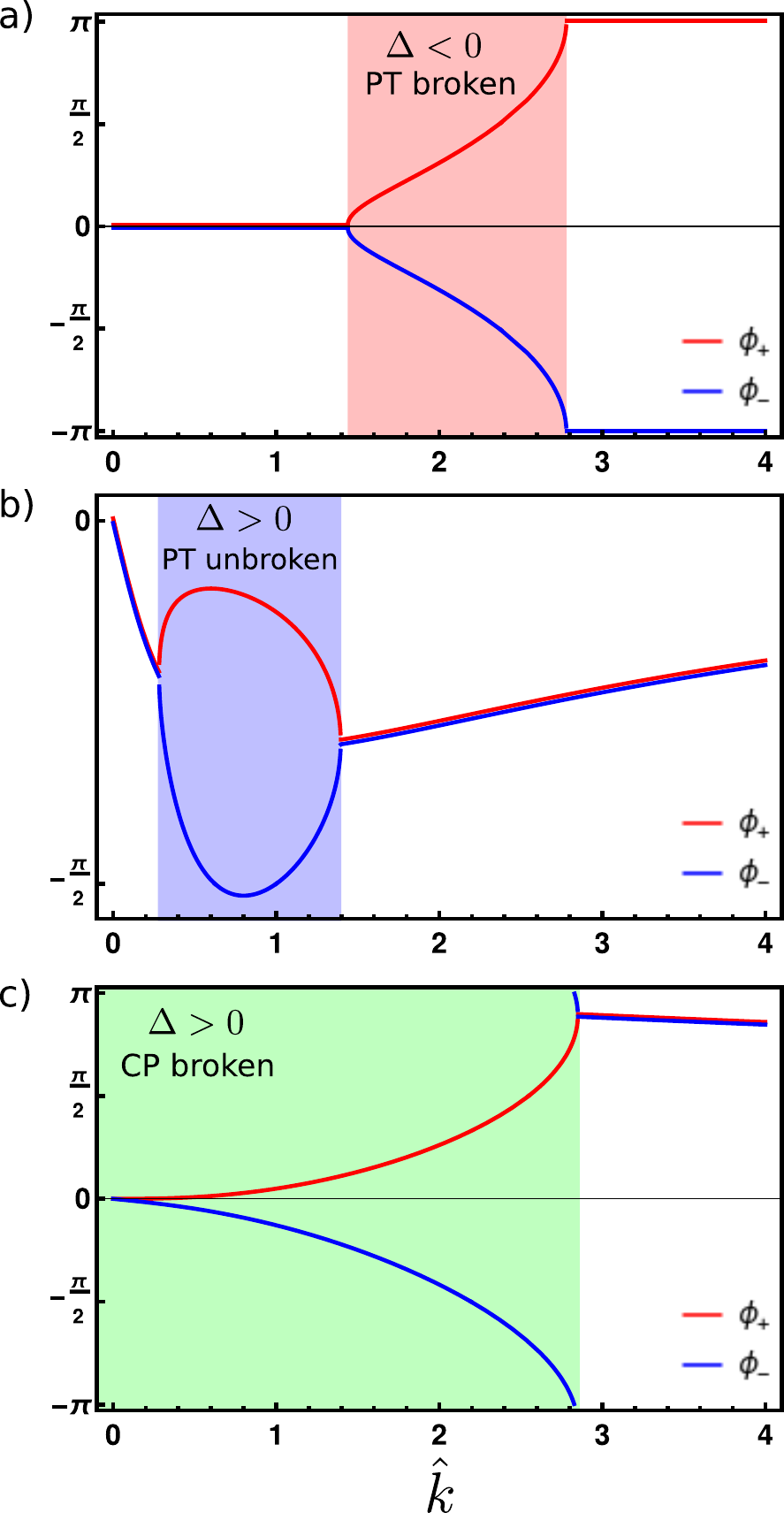}}
\caption{Plots of the phase shift between the layers for each mode as a function of the wavenumber $\hat{k}$. a) Phase shift corresponding to the dispersion relation shown in figure \ref{fig:nor_reg_disp}b where $\hat{l}=0$ and $\gamma = 1$. b) Phase shift corresponding to the dispersion relation shown in figure \ref{fig:noU_reg_disp}b where $\hat{l}=1$ and $\delta = 5$. c) Phase shift corresponding to the dispersion relation shown in figure \ref{fig:nob_reg_disp}b where $\hat{l}=0$ and $\alpha = -1$.}
\label{fig:phasesofeigvecs}
\end{figure}

Finally, it is instructive to notice that, when $\Delta < 0$, the eigenmodes take the form
\begin{equation}
    \bm{\psi_{\pm}} = 
    \begin{pmatrix}
    q_{\pm} e^{i\phi_{\pm}}\\
    1
    \end{pmatrix}
\end{equation}
where $q_{\pm}$ is some real number and $\phi_{\pm}$ is a phase. The eigenvectors, $\bm{\psi_{\pm}}$, determine the relative projection of the instability onto the two layers and also introduce a fixed phase shift between the waves in each layer. Figure \ref{fig:phasesofeigvecs}a shows the dependence of this phase as a function of $k$. We see that in the unbroken regime the phase is zero since the eigenvectors are real while in the broken, or unstable, regime the phase varies from 0 to $\pi$ in between the two exceptional points where the $\Delta = 0$.

Beyond looking at the phase shift between each layer for the unstable mode we can describe the phase difference between the two modes defined by
\begin{equation}
    \Delta \phi = |\phi_+ - \phi_-|.
\end{equation}
 We see in figure \ref{fig:phasesofeigvecs} that $\Delta \phi$ changes from 0 to 2$\pi$ as we go from $\hat{k}=0$ to $k=\infty$. This non trivial winding property is however restricted to the the case $\hat{l}=0$. When $\hat{l}\ne0$ the phase shift is in general not a multiple of $2\pi$.

\subsubsection{No background flow with friction and anti-friction}
\label{sssec:ex_symm_case2}
In this case the system is $P_{xy}P_zT$-symmetric, i.e. it is symmetric by reflection in all three spatial dimensions together with with time-reversal. The corresponding symmetry operator involves layer swapping, and reads
\begin{equation}
    \bm{R_{xy} R_z R_t} =
    \begin{pmatrix}
    0 & -1  \\
    -1 & 0 \\
    \end{pmatrix}.
\end{equation}
The Fourier space representation of the $P_{xy}P_zT$-transformation operator, that acts on the wave operator, is then  obtained by combining this unitary part with complex conjugation, that is $\bm{R_{xy} R_z R_t} \kappa$. The behavior of the eigenvalues of the wave operator, that dictates the stable and unstable regimes, as shown in figure \ref{fig:noU_reg_disp}, thus follows from this  symmetry. 

Again we look at the effect of this symmetry on the eigenmodes. Using equation \eqref{eqn:PT_symm_def}, for a general two-by-two matrix, we have that
\begin{equation}
    \begin{pmatrix}
    c^* & d^* \\
    a^* & b^* 
    \end{pmatrix} =
    \begin{pmatrix}
    b & a \\
    d & c 
    \end{pmatrix}
\end{equation}
which implies that for a two-by-two matrix to be $P_{xy}P_zT$-symmetric we need $d=a^*$ and $c=b^*$, and so, the wave operator can be written as 
\begin{equation}
    \bm{H} = 
    \begin{pmatrix}
    a & b \\
    b^* & a^*
    \end{pmatrix}.
\end{equation}
For this wave operator it can be shown that the eigenvectors can be written as
\begin{equation}
    \bm{v_\pm} = 
    \begin{pmatrix}
    b + i \Im{a} \pm \sqrt{\Delta} \\
    b^* - i \Im{a} \pm \sqrt{\Delta}
    \end{pmatrix},
\end{equation}
which we can see is an eigenvector (unbroken symmetry) of $\bm{R_{xy} R_z R_t}\kappa$
\begin{equation}
    \bm{R_{xy} R_z R_t}\kappa \bm{\psi_{\pm}} = -\bm{\psi_{\pm}}
\end{equation}
when $\Delta > 0$, and is \emph{not} an eigenvector (broken symmetry)
\begin{equation}
    \bm{R_{xy} R_z R_t}\kappa \bm{\psi_{\pm}} \neq \lambda\bm{\psi_{\pm}},
\end{equation}
for any $\lambda$ when $\Delta < 0$. This demonstrates the spontaneous breaking of symmetry for this choice of parameters. Because the eigenvectors are complex values in the when $\Delta > 0$ there is now a phase difference between the solutions in each layer in the unbroken regime. This phase difference is shown in figure \ref{fig:phasesofeigvecs}b, and corresponds to the dispersion relation shown in figure \ref{fig:noU_reg_disp}b.  We see that when the symmetry is unbroken (blue region) the phase between each layer is different for each mode; when the symmetry is broken the phase between the layers becomes the same for each mode, that is, we see a phase locking between the modes.

The phase difference between the modes, $\Delta \phi$, in this case has a more complex behaviour. In this case the $\Delta \phi$ is zero in the \emph{broken} regime. In the unbroken (stable) regime the maximum phase difference varies as a function of non-dimensional parameter, $\delta$, and the wavenumber, $\hat{l}$. 

\subsubsection{No planetary vorticity gradient with friction}
\label{sssec:ex_symm_case3}
In this case we can, by inspection, determine that the system is $P_{xy}P_z$-symmetric, a reflection in the three spatial dimensions, whose corresponding operator thus reads
\begin{equation}
    \bm{R_{xy} R_z} =
    \begin{pmatrix}
    0 & 1  \\
    1 & 0 \\
    \end{pmatrix},
\end{equation}
where we again see the action of layer swapping. The Fourier space representation of the $P_{xy}P_z$-transformation operator, that acts on the wave operator, is then  obtained by combining this unitary part with complex conjugation, that is $\bm{U_{3}} \kappa$. The behavior of the eigenvalues of the wave operator, the suppression of wave propagation, as shown in figure \ref{fig:noU_reg_disp}, thus follows from this  symmetry. 

Now we need to use the definition for CP-symmetry, equation \eqref{eqn:CP_symm_def}, to examine the effect of the operator in this case. For a general two-by-two matrix this implies that
\begin{equation}
    \begin{pmatrix}
    c^* & d^* \\
    a^* & b^* 
    \end{pmatrix} =
    \begin{pmatrix}
    -b & -a \\
    -d & -c 
    \end{pmatrix}
\end{equation}
implying that $d=-a^*$ and $c=-b^*$, and so, the wave operator can be written as 
\begin{equation}
    \bm H =
    \begin{pmatrix}
    a & b \\
    -b^* & -a^*
    \end{pmatrix}.
\end{equation}
For this wave operator the eigenvectors can be written as 
\begin{equation}
    \bm{v_\pm} = 
    \begin{pmatrix}
    b + \Re{a} \pm \sqrt{\Delta} \\
    -b^* -  \Re{a} \pm \sqrt{\Delta}
    \end{pmatrix},
\end{equation}
which is an eigenvector (unbroken symmetry) of $\bm{R_{xy} R_z}\kappa$
\begin{equation}
    \bm{R_{xy} R_z}\kappa \bm{\psi_{\pm}} = -\bm{\psi_{\pm}}
\end{equation}
when $\Delta < 0$ (note that this is opposite to the PT-symmetric examples), and is \emph{not} an eigenvector (broken symmetry)
\begin{equation}
    \bm{R_{xy} R_z}\kappa \bm{\psi_{\pm}} \neq \lambda\bm{\psi_{\pm}},
\end{equation}
for any $\lambda$ when $\Delta > 0$. This demonstrates spontaneous symmetry breaking of $P_{xy}P_z$-symmetry. Again, we can see the behaviour of the phase difference between the the solutions in each layer for the two modes in figure \ref{fig:phasesofeigvecs}c: corresponding to the dispersion relation shown in figure \ref{fig:nob_reg_disp}b.

The phase difference between the modes in this case is zero for $\hat{k} = 0$ becoming equal again at the exceptional point. The phases are equal in the \emph{unbroken} regime. The behaviour changed, however, when $\hat{l} \neq 0$, here, the phase difference $\Delta \phi$ at $\hat{k}=0$ is fixed at $\pi$ changing to 0 at the exceptional point as in the $\hat{l} =0$ example that is shown in figure \ref{fig:phasesofeigvecs}c.

\section{Conclusions}
\label{sec:conclusions}

We have presented a general framework for examining the discrete symmetries of a wide class fluid wave problems, extending this novel area of studies from the initial work of \cite{qinetal2019} and \cite{fuetal2020} which show that Kelvin-Helmholtz instability is an example of spontaneous PT-symmetry breaking. The framework prescribes the method of translating real-space symmetries of partial differential equations into the appropriate Fourier representation of the transformation operators. In doing so, we are now able to immediately write down the relevant anti-unitary operators once a discrete symmetry of a set of partial differential equations has been identified. With these anti-unitary operators we can study the behaviour of the dispersion relation and the eigenmodes of a wave system and look for examples of spontaneous symmetry breaking that shape the spectrum of the flows, including the occurence of instabilities.

In the case of baroclinic instability we were able to show that it is an example of spontaneous PT-symmetry breaking. Further, we showed that the spontaneous symmetry breaking was characterized by the emergence of a phase-shift between waves propagating in each layer within the unstable mode.  In this case we could see that the partial differential equations were symmetric with respect to the transformation $\{x,y,t\} \mapsto \{-x,-y,-t\}$. Using our general framework we were able to show that the Fourier space representation of this transformation is simply complex conjugation. With another choice of parameters we found another example of PT-symmetry, now including the mirror transformation in the vertical direction and we see the spontaneous breaking of this symmetry as the system transitions from stability to instability.

In this study we have been calling the two classes of discrete symmetry  PT- and CP-symmetry (section \ref{ssec:def_of_PTandCP}); these correspond to commutation and anti-commutation relation between the wave operator (or Hamiltonian) with some anti-unitary operator respectively. This follows from the naming used in condensed matter physics where there is a natural means of interpreting charge conjugation \cite{yoshidahatsugai2019,maamachekheniche2020}. As noted in section \ref{sec:introduction}, in the field of optics \cite{antonosyanetal2015,pengetal2016,yangetal2017} the commutation relation is named PT-symmetry while the anti-commutation relation is named anti-PT-symmetry. Our general framework (section \ref{ssec:general}) allows us to understand that the anti-commutation relation, which we have been calling CP-symmetry, is in fact a parity symmetry in the context of layered fluid models. We showed that, for a set of coupled \emph{real-valued} partial differential equations describing a layered fluid model, a pure parity-symmetry is represented in Fourier space as the anti-commutation of the wave operator with some anti-unitary operator which our framework allows us to construct. This result comes from the fact that we are dealing with real-valued partial differential equations and because of this a pure parity transformation for a fluid carries with it the mathematics of charge-conjugation. An example of this symmetry was found for our two-layer model and it was shown that the spontaneous breaking of this symmetry was associated with the suppression of wave propagation.

We expect that this work will open the way to utilizing concepts from the physics of non-Hermitian systems to the realm of geophysical flows, as previously done for electronic and optical systems. This could shine new light on the consequence of discrete symmetries in that context. In turn, geophysical flow model will provide an ideal platform for the development of non-Hermitian physics.  Discrete symmetry breaking also plays a key role in the study of bifurcations and low frequency variability for  turbulent geophysical flows \cite{gallet2012reversals}, including  forced-dissipated climate dynamics  \cite{dijkstrahouches}. Bringing the concept of CP-symmetry could be useful in that context. More generally, by providing a general non-Hermitian framework relating flow instabilities to discrete symmetry breaking, our work  opens the door to a general classification of stable/unstable modes in continuous media, that may appeal  generally to fluids, and in particular to astrophysical and geophysical flows.


\appendix 
\section{Wave operators and dispersion relations}
\label{app:app}
Here we present the  expressions for the wave operators and dispersion relations discussed in section \ref{sec:mode_merging_in_2lqg}. 
\subsection{Case of $r_1 = r_2 = 0$}
\label{app:case_nor}

Here the non-dimensional wave operator is
\begin{equation}
    \bm H = \frac{\hat{k}}{\hat{K}^2(\hat{K}^2 + 8)}
    \begin{pmatrix}
         \hat{K}^4+ -2\gamma (2 \hat{K}^2 + 8)  + 4  & 32 - 16 \gamma \\
         -32 - 16 \gamma & -\hat{K}^4  -2\gamma (2 \hat{K}^2 + 8)  + 4 
    \end{pmatrix}
\end{equation}
where the $\hat{}$ notation denotes dimensionless variables and $\hat{K}^2 = \hat{k}^2 + \hat{l}^2$. The dispersion relation is

\begin{equation}
\label{eqn:case1_disprel}
    \omega_{\pm} = \frac{\hat{k}}{\hat{K}^2 \left(\hat{K}^2+8\right)} \left(-4 \gamma 
   \left(\hat{K}^2+4\right) \pm \sqrt{256 \gamma ^2+\hat{K}^4\left(\hat{K}^4-64\right)} \right).
\end{equation}

\subsection{Case of $U=0$ and $r_1=-r_2=r$}
\label{app:case_noU}
Here the wave operator for the problem becomes

\begin{equation}
    \bm H = 
    \frac{i}{\hat{K}^2 \left(\hat{K}^2+8\right)}\begin{pmatrix}
     \left(\hat{K}^2+4\right) \left(\hat{k}^2+i \delta  \hat{k}+\hat{l}^2\right) & -4 \left(\hat{k}^2-i \delta 
   \hat{k}+\hat{l}^2\right) \\
 4  \left(\hat{k}^2+i \delta  \hat{k}+\hat{l}^2\right) & - \left(\hat{K}^2+4\right) \left(\hat{k}^2-i \delta 
   \hat{k}+\hat{l}^2\right) \\
    \end{pmatrix},
\end{equation}

and the dispersion relation is

\begin{equation}
    \omega_{\pm} =\frac{-\delta  \hat{k} \left(\hat{K}^2+4\right) \pm \sqrt{4 \hat{k}^2 \left(4 \delta ^2-\hat{l}^4 \left(\hat{l}^2+6\right)\right)-4 \hat{k}^6 \left(\hat{l}^2+2\right)-6 \hat{k}^4 \hat{l}^2 \left(\hat{l}^2+4\right)-\hat{k}^8-\hat{l}^6
   \left(\hat{l}^2+8\right)}}{\hat{K}^2 \left(\hat{K}^2+8\right)}.
\end{equation}

\subsection{Case of $\beta = 0$ and $r_1=r_2=r$}
\label{app:case_nob}
Now the wave operator becomes 

\begin{equation}
    \bm H = 
    \begin{pmatrix}
    i \alpha  \left(1-\frac{4}{\hat{K}^2+8}\right)+\hat{k} \left(-\frac{4}{\hat{K}^2+8}+1-\frac{4}{\hat{K}^2}\right) & \frac{4 i \left(\alpha  \hat{l}^2+\hat{k} \left(\alpha 
   \hat{k}-8 i\right)\right)}{\left(\hat{K}^2\right) \left(\hat{K}^2+8\right)} \\
 \frac{4 i \left(\alpha  \hat{l}^2+\hat{k} \left(\alpha  \hat{k}+8 i\right)\right)}{\left(\hat{K}^2\right) \left(\hat{K}^2+8\right)} & i \alpha 
   \left(1-\frac{4}{\hat{K}^2+8}\right)+\hat{k} \left(\frac{4}{\hat{K}^2+8}-1+\frac{4}{\hat{K}^2}\right) \\
    \end{pmatrix}
\end{equation}

and the dispersion relation is 

\begin{equation}
    \omega_{\pm} = \frac{i \alpha 
   \left(\hat{K}^2+4\right)\pm \sqrt{ \left(\hat{k}^2 \left(\hat{K}^4-64\right)-16 \alpha ^2\right)}}{ \left(\hat{K}^2+8\right)}.
\end{equation}

\section{Two-dimensional parity in Fourier-space}

\label{app:proof}

Consider the general layered and linearized fluid model
\begin{equation}
    \partial_t \bm \psi(x,y) = \mathcal{L}_{\bm \lambda} \left[\partial_x,\partial_y \right]  \bm \psi(x,y), 
\end{equation}
discussed in section \ref{ssec:classical_to_fourier}. The operator, $\mathcal{L}$ can be either a differential operator or a pseudo-differential operator.  When $\mathcal{L}$ is a pseudo-differential operator, \eqref{eqn:general} can be written as $\partial_t \mathcal{A} \bm \psi = \mathcal{D} \bm \psi$ or alternatively, $\partial_t \bm \psi = \mathcal{A}^{-1} \mathcal{D} \bm \psi$, where $\mathcal{A}$ and $\mathcal{D}$ are differential operators. In Fourier space the operator $\mathcal{L}_{\bm \lambda}$ becomes the wave operator, $\bm H$, which is a function of $k$ and $l$. In general we can write the matrix elements of the waves operator, $h_{\mu\nu}$ as
\begin{equation}
    h_{\mu\nu} = \frac{iP^{\mu\nu}_{n,m}(ik,il)}{Q_{\tilde{n},\tilde{m}}(ik,il)},
\end{equation}
where $P$ is a polynomial in the two variables, $(ik,il)$, of degree $(n,m)$ respectively, such that 
\begin{equation}
    i P^{\mu\nu}_{n,m}(ik,il) = i \sum_{\alpha,\beta} a_{\alpha,\beta} (ik)^\alpha (il)^\beta,
\end{equation}
and $Q$ is a polynomial in the two variables, $(ik,il)$, of degree $(\tilde{n},\tilde{m})$ respectively, such that
\begin{equation}
    Q_{\tilde{n},\tilde{m}}(ik,il) = \sum_{\alpha,\beta} b_{\alpha,\beta} (ik)^\alpha (il)^\beta.
\end{equation}
In the case that $\mathcal{L}$ is simply a differential operator $Q_{\tilde{n},\tilde{m}}=1$; in the case that $\mathcal{L}$ is a pseudo-differential operator $Q_{\tilde{n},\tilde{m}}$ is a polynomial associated with the inverted differential operator (this would be determinant of the Fourier representation of the differential operator $\mathcal{A}$). Each term of the polynomials $\{iP^{\mu\nu}_{n,m}\}$ and $Q_{\tilde{n},\tilde{m}}$ is either real or imaginary depending on the value of the powers $\alpha$ and $\beta$. We define the transformation 
\begin{equation}
 F:\, (k,l) \mapsto (-k,-l), 
\end{equation}
 and consider its effect on the polynomials in each of  the following cases.
\begin{itemize}
    \item The even terms of $\{iP^{\mu\nu}_{n,m}\}$, when $\alpha+\beta$ is an even number; the sign of these terms are not changed by the transformation, $F$, and are imaginary. Thus, we can see that for these terms the transformation $F$ is equivalent to taking the complex conjugate and multiplying by -1.
    \item The odd terms of $\{iP^{\mu\nu}_{n,m}\}$, when $\alpha+\beta$ is an odd number; the sign of these terms are changed by the transformation, $F$, and are real. Thus, we can see that for these terms the transformation $F$ is equivalent to taking the complex conjugate and multiplying by -1.
    \item The even terms of $Q_{\tilde{n},\tilde{m}}$, when $\alpha+\beta$ is an even number; the sign of these terms are not changed by the transformation, $F$, and are real. Thus, we can see that for these terms the transformation $F$ is equivalent to taking the complex conjugate.
    \item The even terms of $Q_{\tilde{n},\tilde{m}}$, when $\alpha+\beta$ is an even number; the sign of these terms are changed by the transformation, $F$, and are imaginary. Thus, we can see that for these terms the transformation $F$ is equivalent to taking the complex conjugate.
\end{itemize}
Putting this together we see that all the elements of $\bm H$ transform under $F$ as $h_{\mu\nu}(k,l) \mapsto h_{\mu\nu}(-k,-l) = -h_{\mu\nu}^*(k,l)$ and we can therefore write that
\begin{equation}
    \label{eqn:PTeqiv}
    \bm{H}_{\bm \lambda}[-k,-l] = -\bm{H}^*_{\bm \lambda}[k,l].
\end{equation}


\bibliography{./tdbib}


%
%

%



\end{document}